\documentclass[%
reprint,
superscriptaddress,
amsmath,
amssymb,
aps,
pra,
]{revtex4-2}

\usepackage{graphicx}
\usepackage{dcolumn}
\usepackage{bm}
\usepackage{siunitx}
\usepackage{color}
\usepackage{array}
\usepackage{booktabs}
\usepackage{lineno}
\newcommand{\GHz}{\mathrm{GHz}}

\definecolor{dg}{RGB}{0, 0, 200}
\usepackage{hyperref}
\usepackage[nameinlink]{cleveref}

\crefname{figure}{Fig.}{Figs.}
\Crefname{figure}{Fig.}{Figs.}

\crefname{equation}{Eq.}{Eqs.}
\Crefname{equation}{Eq.}{Eqs.}

\begin{document}
\preprint{APS/123-QED}

\title{90\%-Efficient Optical Storage in a Rare-Earth Crystal with a Low-Loss Impedance-Matched Cavity}

\author{Weiye Sun}

\affiliation{%
Southern University of Science and Technology, Shenzhen 518055, China
}%
\affiliation{%
International Quantum Academy, and Shenzhen Branch,
Hefei National Laboratory, Shenzhen, 518048, China
}%
\affiliation{These authors contributed equally to this work.}%
\author{Fudong Wang}%
\email{fdwang.phys@foxmail.com}

\affiliation{%
International Quantum Academy, and Shenzhen Branch,
Hefei National Laboratory, Shenzhen, 518048, China
}%
\affiliation{These authors contributed equally to this work.}%

\author{Mucheng Guo}%
\affiliation{%
Southern University of Science and Technology, Shenzhen 518055, China
}%
\affiliation{%
International Quantum Academy, and Shenzhen Branch,
Hefei National Laboratory, Shenzhen, 518048, China
}%
\author{Xiantong An}%
\affiliation{%
Southern University of Science and Technology, Shenzhen 518055, China
}%
\author{Zhenqi Xu}%
\author{Zhehao Xu}%
\author{Xingmin He}
\affiliation{%
Southern University of Science and Technology, Shenzhen 518055, China
}%
\affiliation{%
International Quantum Academy, and Shenzhen Branch,
Hefei National Laboratory, Shenzhen, 518048, China
}%

\author{Matthew J. Sellars}%
\affiliation{%
Centre of Excellence for Quantum Computation and Communication Technology,
Research School of Physics, Australian National University, Canberra, ACT 0200, Australia
}%

\author{Shuping Liu}%
 \email{liushuping@iqasz.cn}

\author{Manjin Zhong}
 \email{manjin.zhong@gmail.com}
\affiliation{%
International Quantum Academy, and Shenzhen Branch,
Hefei National Laboratory, Shenzhen, 518048, China
}%


%

\date{\today}

\begin{abstract}

High-efficiency optical storage in rare-earth-ion-doped crystals is challenging because of their typically weak optical absorption, which makes cavity-enhanced memories highly sensitive to parasitic intracavity loss. Here we demonstrate a 90.1(5)\% storage efficiency for coherent optical pulses in an impedance-matched cavity atomic-frequency-comb memory based on $\mathrm{Eu}^{3+}$:$\mathrm{Y}_2\mathrm{SiO}_5$. The Brewster-angle configuration of the crystal and fully cryogenic Fabry–P\'erot cavity reduce the round-trip intracavity loss to 0.37\%. 
Quantitative modeling shows that further improvement is limited
primarily by AFC preparation and its associated dispersive response rather than by residual cavity loss, identifying the principal requirements for further approaching unity efficiency.

\end{abstract}

\maketitle


Optical quantum memories bridge flying photonic qubits and stationary quantum systems, providing a key interface for quantum networking and quantum information processing \cite{Kimble2008, RevModPhys.83.33, RevModPhys.79.135, PhysRevLett.81.5932, Lago-Rivera2021, Liu2024,PhysRevA.64.010301,PhysRevLett.98.190503,PhysRevLett.110.133601, Liu202410}. Rare-earth-ion-doped (REI-doped) crystals are promising platforms for quantum memories, combining long optical and spin coherence times with broad inhomogeneous linewidths to enable long-duration storage and high-capacity multimode operation \cite{PhysRevLett.120.183602,Wang2019,Cho16,Hosseini2011,Guo2019,kbwj-md9n,ZhongM2015,PRXQuantum.6.010302}. Among various quantum memory protocols, atomic frequency comb (AFC) storage is especially well suited to these materials, offering intrinsic temporal multimode capacity and compatibility with spin-wave storage for on-demand retrieval \cite{PhysRevB.68.085109,PhysRevLett.95.030506,PhysRevB.73.075101, PhysRevB.101.184430, Tiranov2026,PhysRevB.85.115111, PhysRevB.70.214116}.  Achieving high AFC efficiency, however, remains challenging in weakly absorbing REI-doped crystals, while reabsorption during forward recall limits the ideal free-space AFC efficiency to 54\% \cite{PhysRevA.79.052329,PhysRevA.81.033803,ZangAllen_2025}.

Embedding the REI-doped crystal in an asymmetric Fabry–P\'erot (FP) cavity provides a route around these limitations. Under impedance matching condition, the incident field can be efficiently absorbed even at small single-pass optical depth, allowing the storage efficiency to approach unity in the absence of parasitic loss  \cite{TittelWolfgang_2025}. This approach was first demonstrated for an AFC memory in \qty{500}{ppm} $\mathrm{Pr}^{3+}$:$\mathrm{Y}_2\mathrm{SiO}_5$ ($\mathrm{Pr}^{3+}$:YSO), where a crystal with only 10\% single-pass absorption yielded a storage-and-retrieval efficiency of 56\% \cite{PhysRevLett.110.133604}.

In the weak-absorption regime, however, impedance matching becomes particularly sensitive to parasitic intracavity loss. When the parasitic loss becomes comparable to the effective AFC absorption, an increasing fraction of the intracavity field is dissipated without contributing to storage, substantially reducing the attainable efficiency.  Recent cavity-enhanced $\mathrm{Eu}^{3+}$:YSO memories have reached storage efficiencies of 80.3(7)\% for weak coherent pulses \cite{Meng2026}, demonstrating the potential of impedance-matched cavities while placing increasingly stringent requirements on parasitic loss and AFC preparation.

Here we demonstrate 90.1(5)\% storage efficiency for coherent optical pulses in an impedance-matched cavity AFC memory based on $\mathrm{Eu}^{3+}$:YSO. By placing the entire FP cavity inside a cryogenic environment and using a Brewster-angle crystal configuration, we reduce the round-trip intracavity loss to 0.37\%. Quantitative modeling shows that further reduction of parasitic cavity loss offers diminishing returns at this loss level, shifting the main opportunity for higher efficiency toward improved AFC preparation and control of its dispersive response.

\begin{figure*}
\includegraphics{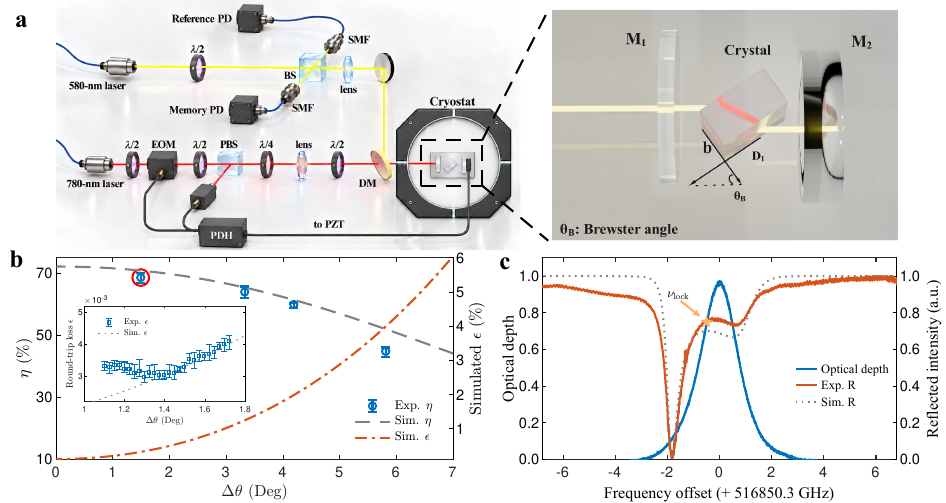}
\caption{\label{fig1} \textbf{a}, Schematic of the experimental setup. A stabilized 780-nm laser is used for cavity locking, while a 580-nm laser is used for optical storage. The AFC preparation light and the input signal to be stored are coupled into the cavity along the same optical path. The enlarged view shows the storage cavity, with $\mathrm{Eu}^{3+}$:YSO crystal placed between a plane mirror (M$_1$) and a concave mirror (M$_2$). The Brewster angle $\theta_{\mathrm{B}}$ is indicated.
\textbf{b}, Measured storage efficiency (blue open circles), simulated storage efficiency (gray dashed curve), and simulated round-trip intracavity loss $\epsilon$ (orange dash-dotted curve) as functions of the angular deviation $\Delta\theta$ from the Brewster angle. The simulations assume impedance matching and an AFC finesse of 5. The simulated storage efficiency is calculated using \cref{eq:eta_AFC}. The red circle marks the operating point corresponding to the minimum loss in the inset, which is used in the subsequent memory experiments. The inset shows the measured round-trip intracavity loss during fine angular adjustment using a micrometer head (blue open squares) and the corresponding simulated loss (gray dotted curve).
\textbf{c}, Measured optical depth (blue curve) and measured cavity reflection spectrum (orange curve) as functions of the frequency offset relative to \qty{516.8503}{\THz}. The gray dotted curve shows the simulated reflection spectrum, and the arrow indicates the cavity-locking frequency $\nu_{\mathrm{lock}}$. The measured reflection minimum occurs at a frequency offset of approximately \qty{-1.8}{\GHz}.}
\end{figure*}

As shown in \cref{fig1}(a), the memory consists of a $\mathrm{Eu}^{3+}$:YSO crystal placed inside an asymmetric FP cavity, with the entire assembly housed in a 3-K cryostat. The crystal contains naturally abundant Eu$^{3+}$ ions at a concentration of \qty{1000}{ppm} and has dimensions of \qtyproduct{5.0 x 5.5 x 3.0}{\mm} $(D_1 \times D_2 \times b)$. The cavity consists of a plane input mirror $\mathrm{M}_1$ with reflectivity $R_1 = 0.8$ and a highly reflective concave mirror $\mathrm{M}_2$ with $R_2 > 0.9995$ at \qty{580}{\nm} and \qty{3}{\K}. The crystal is oriented at the Brewster angle, with the incident polarization parallel to the $D_1$-$b$ plane to suppress reflection from the uncoated crystal surfaces while addressing the strongly absorbing polarization. A 780-nm laser is used for cavity locking, while the 580-nm memory optical field addresses the ${}^{7}F_0-{}^{5}D_0$ transition of $\mathrm{Eu}^{3+}$ at site 1 \cite{PhysRevB.68.085109, ZhongM2015,PRXQuantum.6.010302}.

\begin{figure*}
\includegraphics{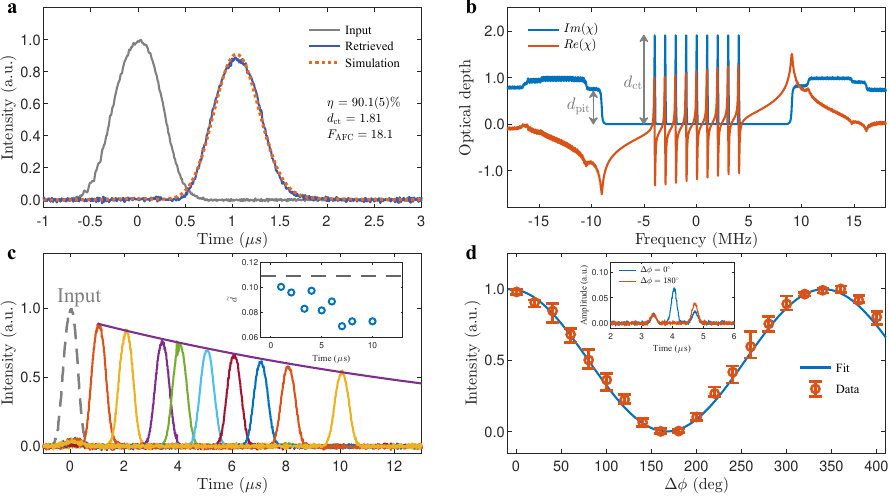}
\caption{\label{fig2} Storage efficiency and phase coherence of the optical memory.
\textbf{a}, Normalized input pulse (gray solid curve) and retrieved echo (blue solid curve) for an AFC tooth spacing of $\Delta = \qty{1}{\MHz}$. The calibrated storage efficiency is $\eta = 90.1(5)\%$. The center of the retrieved echo is delayed by approximately \qty{0.05}{\us} relative to the nominal recall time of $1/\Delta = \qty{1}{\us}$. The orange dotted curve shows the simulated echo, with a predicted storage efficiency of 91.8\%.
\textbf{b}, Simulated AFC spectrum for $\Delta = \qty{1}{\MHz}$ and a finesse of $F_{\mathrm{AFC}} = 18.1$. The blue and orange curves show the imaginary and real parts of the optical susceptibility $\chi$, respectively. The gray arrows indicate the optical depths $d_{\mathrm{ct}}$ and $d_{\mathrm{pit}}$.
\textbf{c}, Measured retrieved echoes for nine different storage times. The gray dashed curve shows the input pulse, and the purple solid curve shows an exponential decay fit. The inset shows the simulated values of $\widetilde{d}$ as a function of storage time (blue open circles). The gray dashed line indicates the impedance matching condition, $\widetilde{d}\simeq0.11$.
\textbf{d}, Interference fringe as a function of the relative phase $\Delta\phi$ between the time-bin modes. The inset shows the temporal interference traces for $\Delta\phi=\qty{0}{\degree}$ (blue) and $\Delta\phi=\qty{180}{\degree}$ (orange).}
\end{figure*}

For the cavity-enhanced AFC memory, the relevant absorption is the effective AFC optical depth $\widetilde{d} = d_\mathrm{ct}/F_\mathrm{AFC}$, where $d_\mathrm{ct}$ is the optical depth of the comb teeth and $F_\mathrm{AFC} = \Delta/\gamma$ is the AFC finesse. Including a round-trip parasitic intracavity loss $\epsilon$, the impedance matching condition becomes \cite{PhysRevA.82.022310}
\[
R_1 = R_2e^{-2\widetilde{d} - \epsilon}.
\]
For the measured cavity parameters, the impedance matching condition requires $\widetilde{d} \simeq 0.11$. If $\epsilon\ll\widetilde{d}\ll 1$, the storage efficiency under impedance matching is approximately \cite{PhysRevA.82.022311, PhysRevA.82.022310,Jobez_2014,Meng2026}
\begin{equation}
\eta=\frac{\eta_\mathrm{c}\cdot\eta_\mathrm{deph}}{(1+\frac{\epsilon}{4\widetilde{d}})^4},
\label{eq:eta_AFC}
\end{equation}
where $\eta_\mathrm{c}$ is the spatial mode-matching efficiency and $\eta_\mathrm{deph}$ describes dephasing due to the finite AFC finesse; for Gaussian comb teeth, $\eta_{\mathrm{deph}}=exp[-\pi^{2}/(2\mathrm{ln}2F_{\mathrm{AFC}}^{2})]$ \cite{PhysRevA.79.052329}. \cref{eq:eta_AFC} shows that, when $\widetilde{d}$ is small, parasitic loss must remain well below the effective AFC absorption to avoid a substantial efficiency penalty.

We therefore designed the cavity to minimize parasitic loss within the resonator. Placing the entire FP cavity inside the cryostat eliminates cryostat-window losses from the cavity round trip, yielding an empty-cavity round-trip loss of 0.12\%. Reflection from the uncoated crystal surfaces is suppressed by operating at the Brewster angle. As shown by the simulation results in \cref{fig1}(b), a \qty{3}{\degree} deviation from the Brewster angle introduces approximately 1\% additional round-trip loss, corresponding to a reduction in storage efficiency of about 5.5 percentage points. Fine angular adjustment reduces the crystal-related reflection loss to approximately 0.2\%, resulting in a total round-trip intracavity loss of 0.32\% at room temperature and 0.37\% at 3 K.

\cref{fig1}(c) shows the cavity reflection together with the crystal optical-depth spectrum before AFC preparation. The reflection reaches a minimum at a detuning of \qty{-1.8}{\GHz} from the absorption center, where the crystal optical depth is approximately 0.11. The reflection minimum corresponds to 99.4\% cavity coupling efficiency, in good agreement with the calculated reflection spectrum (see Supplemental Materials for details).

We next optimize the AFC parameters under the impedance matching constraint. Because $\widetilde{d} = d_\mathrm{ct}/F_\mathrm{AFC}$, increasing the comb-tooth optical depth $d_\mathrm{ct}$ permits operation at higher $F_{\mathrm{AFC}}$ while maintaining the required effective optical depth, thereby reducing finite-finesse dephasing. At the operating frequency \(\nu_\mathrm{lock} = \qty{-0.46}{\GHz}\), the initial optical depth is $d_\mathrm{ct}$ = 0.75. An absorption-enhancement sequence \cite{PhysRevLett.128.180501, Meng2026} increases $d_\mathrm{ct}$ to $1.91\pm0.2$, allowing $F_{\mathrm{AFC}} \simeq 17.4$ while maintaining maintaining $\widetilde{d}\simeq0.11$.

Under these conditions, we obtain a maximum storage efficiency of 90.1(5)\% for a Gaussian input pulse with a full width at half maximum (FWHM) of  \qty{0.53}{\us} and a storage time of  \qty{1}{\us} [\cref{fig2}(a)]. The negligible prompt leakage provides an additional signature of operation close to the impedance matching condition.

\begin{figure*}
\includegraphics{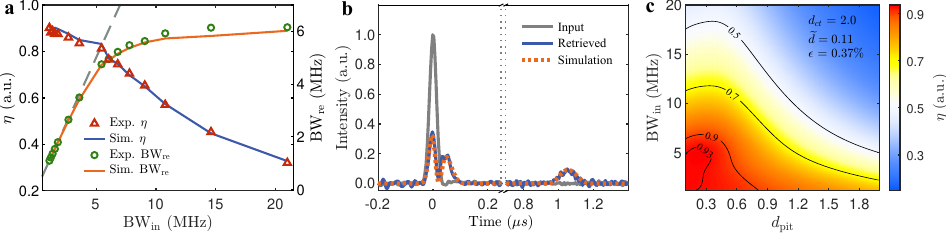}
\caption{\label{fig3} Bandwidth dependence of the optical memory.
\textbf{a}, Storage efficiency $\eta$ (left axis) and retrieved-echo bandwidth $\mathrm{BW}_{\mathrm{re}}$ (right axis) as functions of the input-pulse bandwidth $\mathrm{BW}_{\mathrm{in}}$. As the input pulse becomes wider, the $\mathrm{BW}_{\mathrm{re}}$ (green circles) deviates from the linear region (gray dashed line), and $\eta$ (red triangles) drops rapidly when the input pulse width exceeds \qty{2}{\MHz}. Simulated results are also given for $\eta$ (blue solid line) and $\mathrm{BW}_{\mathrm{re}}$ (orange solid line). \textbf{b}, Measured temporal profiles of the input pulse (gray solid curve) and the leakage light and retrieved echo (blue solid curve) for an input-pulse bandwidth of $\mathrm{BW}_{\mathrm{in}}=\qty{21}{\MHz}$. The orange dotted curve shows the simulated temporal profile, which approximately reproduces the peak intensity and temporal width of the retrieved echo, as well as the shape and peak intensity of the leakage light. \textbf{c}, Simulated storage efficiency as a function of $d_{\mathrm{pit}}$ and $\mathrm{BW}_{\mathrm{in}}$, calculated for $d_{\mathrm{ct}}=2.0$, a round-trip intracavity loss of $\epsilon=0.37\%$, and the impedance matching condition $\widetilde{d}=0.11$. The color scale represents $\eta$.} 
\end{figure*}

To identify the remaining efficiency limitations, we model the complete cavity-memory response using the prepared AFC absorption profile and the corresponding dispersion obtained from the Kramers–Kronig relations \cite{Jobez_2014,Taherizadegan_2024, Meng2026}. The optimized model yields $d_\mathrm{ct} = 1.81$, $F_{\mathrm{AFC}} = 18.1$, and a storage efficiency of 91.8\%, close to the measured value. By comparison, \cref{eq:eta_AFC} predicts an efficiency of 93.9\% when evaluated using the measured cavity parameters and Gaussian-shaped comb teeth. Using the fitted rather than measured $d_\mathrm{ct}$ changes the predicted efficiency by only about 0.2 percentage points, while dispersion associated with the finite spectral pit reduces the predicted efficiency by approximately 1.8 percentage points. This dispersion introduces a frequency-dependent cavity response and produces a group delay of approximately  \qty{0.05}{\us}, consistent with the observed displacement of the AFC echo [\cref{fig2}(a)]. These results show that, at the present round-trip loss of 0.37\%, further improvement is limited primarily by AFC preparation and its associated dispersion rather than by residual intracavity loss.

We next investigate the storage-time dependence [\cref{fig2}(c)]. The efficiency remains above 50\% at a storage time of \qty{10}{\us}, and an exponential fit yields an effective decay time of \qty{35.8(8)}{\us}. At longer storage times, the reduced comb spacing makes the AFC increasingly sensitive to preparation imperfections and residual spectral structure. Simulations constrained by the measured AFC echoes show that $\widetilde{d}$ progressively departs from the impedance-matched value as the storage time increases, resulting in increased prompt leakage and reduced storage efficiency. These results identify AFC-preparation imperfections and the resulting departure from impedance matching as a major contribution to the observed efficiency decay.

We then characterize the phase coherence of the memory using a double-AFC structure as an unbalanced interferometric analyzer \cite{PhysRevLett.125.260504}. Time-bin states modes $|\psi\rangle = |E\rangle + e^{i\Delta\phi}|L\rangle$ are stored using two AFCs with storage times of \qty{3.3}{\us} and \qty{4.0}{\us}, chosen such that the corresponding echoes overlap and interfere. Varying $\Delta\phi$ produces the fringes shown in \cref{fig2}(d), with a visibility of V = $(99.6\pm 2.1)\%$, corresponding to an interference-based fidelity of F = (V + 1)/2 = $(99.8\pm 1.1)\%$ \cite{MaYu2021}.

Finally, we examine the effect of the AFC dispersive response on the memory bandwidth. With the AFC bandwidth of \qty{9}{ \MHz} and spectral-pit width fixed, the input bandwidth $\mathrm{BW}_{\mathrm{in}}$ is varied from 1.1 to \qty{21}{ \MHz} at a storage time of \qty{1}{ \us}. As $\mathrm{BW}_{\mathrm{in}}$ increases, the storage efficiency decreases and the retrieved bandwidth departs progressively from the linear response, approaching approximately \qty{6}{ \MHz} once $\mathrm{BW}_{\mathrm{in}}$ exceeds the \qty{9}{ \MHz} AFC bandwidth [\cref{fig3}(a)]. These measurements can also be quantitatively explained with numerical simulations. For example, the numerical model reproduces both the retrieved echo and the prompt leakage when $\mathrm{BW}_{\mathrm{in}}$ = \qty{21}{ \MHz} [\cref{fig3}(b)].

\cref{fig3}(c) further isolates the influence of the residual optical depth $d_{\mathrm{pit}}$ at the edge of the spectral pit. For $d_{\mathrm{pit}}$ between 0.1 and 0.3, the storage efficiency remains above 50\% for input bandwidths up to approximately 17–\qty{18}{ \MHz}, and can exceed 93\% for $\mathrm{BW}_{\mathrm{in}} <$ \qty{6}{ \MHz}. As $d_{\mathrm{pit}}$ increases, both the usable bandwidth and the maximum storage efficiency decrease. These results show that residual absorption surrounding the AFC limits both efficiency and bandwidth through its associated dispersive response.

At the measured round-trip intracavity loss of 0.37\%, even complete elimination of the residual crystal-surface reflection is expected to increase the storage efficiency by only about 1.5 percentage points. The dominant opportunity for further improvement therefore lies in the prepared AFC, particularly finite-finesse dephasing and dispersion associated with the surrounding spectral structure.

Because the effective optical depth required for impedance matching is fixed by the cavity parameters, increasing $d_{\mathrm{ct}}$ would permit operation at higher $F_{\mathrm{AFC}}$ while maintaining the required $\tilde d$, thereby reducing finite-finesse dephasing. The challenge is to achieve this increase without introducing additional residual absorption or spectral structure around the AFC, as these features can induce a dispersive response that limits both efficiency and bandwidth. Narrower-linewidth preparation lasers, improved crystal optical quality, and optimized spectral-preparation protocols therefore provide direct routes toward higher efficiency.

More generally, these results demonstrate that weak optical absorption does not preclude high-efficiency storage when parasitic cavity loss is kept well below the effective memory absorption. Extending this low-loss impedance-matched architecture to spin-wave AFC storage could provide a route to combining high optical efficiency with on-demand, long-lived storage.

\section*{Acknowledgments}
We thank （Dr. Zongquan Zhou, Dr. Ming Jin, Dr. Tianxiang Zhu） for valuable discussions. This work was supported by the Quantum Science and Technology-National Science and Technology Major Project (QNMP, Grant No. 2021ZD0301204), the National Natural Science Foundation of China (Grant No. 12304454, 12004168, 11904159), National Key Research and Development Program of China (Grant No. 2022YFB3605800), Guangdong Basic and Applied Basic Research Foundation (Grant No. 2021A1515110191), Guangdong Innovative and Entrepreneurial Research Team Program (Grant No. 2019ZT08X324), the Key-Area Research and Development Program of Guangdong Province (Grant No. 2018B030326001), and The Science, Technology and Innovation Commission of Shenzhen Municipality (KQTD202108110900-49034).
\bibliography{apssamp}

\nolinenumbers
\end{document}